\documentstyle[12pt]{article}
\newcommand{\beq}{\begin{equation}}
\newcommand{\eeq}{\end{equation}}
\newcommand{\bem}{\begin{displaymath}}
\newcommand{\eem}{\end{displaymath}}
\newcommand{\bef}{\begin{fig}}
\newcommand{\eef}{\end{fig}}
\newcommand{\ber}{\begin{bref}}
\newcommand{\eer}{\end{bref}}
\newcommand{\beqa}{\begin{eqnarray}}
\newcommand{\nnq}{\nonumber}
\newcommand{\eeqa}{\end{eqnarray}}
\newcommand{\labeleq}[1]{\label{#1}}
\newcommand{\labeleqa}[1]{\label{#1}}
\newcommand{\req}[1]{(\ref{#1})}
\newcommand{\bm}[1]{\mbox{\boldmath{$#1$}}}
\newcommand{\sech}{\mbox{$\,$sech$\,$}}
\newcommand{\citr}[1]{${^{\cite{#1}}}$}
\newcommand{\doublespace}{\baselineskip 8.4667mm \parskip 0.0mm}

\newenvironment{fig}{\refstepcounter{nfig}
    \noindent \small {\bf Figure \thenfig. } \ }{\vskip10pt \par }
\def\barr{\hbox{I\hskip -.5ex R}}

\def\sech{{\hbox { sech}}}

\def\p{ \partial}
\def\ov{ \over}

\def\R{{\cal R}}

\def\tanh{{\hbox {tanh}}}
\def\ep{\epsilon}
\def\al{\alpha}

\def\ka{\kappa}
\def\la{\lambda}

\def\ir{\int^{+\infty}_{-\infty}}

\def\haf{{1 \ov 2}}
\def\ze{\zeta}
\def\de{\delta}
\def\Del{\Delta}
\def\ga{\gamma}
\def\la{\lambda}

\def\avi{\int^{\zeta}_{-h}}
\def\avi0{\int^{\zeta}_{-h_0}}

\def\K{{\cal{K}}}

\def\Ham{Hamiltonian \hskip 2pt}

\def\GN{Green-Naghdi \hskip2pt}

\def\KdV{Korteweg-de Vries \hskip2pt}

\begin{document}
\doublespace
%
\title{ An integrable shallow\\
water equation with peaked solitons}
\author{Roberto Camassa and
Darryl D. Holm \\[12pt]
\small
Theoretical Division and Center for Nonlinear Studies\\
\small
Los Alamos National Laboratory\\
\small
Los Alamos, NM\ \ 87545\\[12pt]
}
\maketitle
\vskip1.5in
\begin{abstract}
\noindent
We derive a new completely integrable dispersive shallow water equation that is
biHamiltonian and thus possesses an infinite number of conservation laws in
involution.
The equation is obtained by using an asymptotic expansion directly in the
Hamiltonian for Euler's equations in the shallow water regime.
The soliton solution for this equation has a limiting form
that has a discontinuity in the first
derivative at its peak.

\vspace{0.5cm}

\vspace{1.5cm}
\footnoterule
\noindent
{\it Submitted to Phys. Rev. Lett.}
\end{abstract}

\clearpage

%
Completely integrable nonlinear partial differential equations arise at various
levels of
approximation in shallow water theory. Such equations possess soliton solutions
- coherent
(spatially localized) structures that interact nonlinearly among themselves
then re-emerge,
retaining their identity and showing particle-like scattering behavior.
In this paper, we use Hamiltonian methods to derive a new completely integrable
dispersive shallow water equation,
\beq
u_t+\ka u_x -u_{xxt}+3uu_x=2u_xu_{xx}+uu_{xxx}\ ,
\labeleq{cluv}
\eeq
where $u$ is the fluid velocity in the $x$ direction (or equivalently the
height of
the water's free surface above a flat bottom), $\ka$ is a constant
related to the critical shallow water wave speed, and subscripts denote partial
derivatives.
This equation retains higher order terms (the right-hand side) in a small
amplitude
expansion of incompressible Euler's equations for unidirectional motion of
waves at the
free surface under the influence of gravity.
Dropping these terms leads to the Benjamin-Bona-Mahoney (BBM) equation,
or at the same order, the \KdV (KdV) equation.
Our extension of the BBM equation possesses soliton solutions whose limiting
form as $\ka \rightarrow 0$ have peaks where first derivatives are
discontinuous.
These ``peakons" dominate the solution of the initial value problem for this
equation with
$\ka=0$. The way a smooth initial condition breaks up into a train of peakons
is by developing
a verticality at each inflection point with negative slope, from which a
derivative
discontinuity emerges.
Remarkably, the multisoliton solution
is obtained by simply superimposing the single peakon solutions
and solving for the evolution of their amplitudes and the positions of their
peaks as a completely integrable finite dimensional \Ham system.

Our equation is biHamiltonian, i.e., it can be expressed in \Ham form in two
different ways.
The ratio of its two (compatible) \Ham operators is a recursion operator that
produces an
infinite sequence of conservation laws. This biHamiltonian property is used to
recast
our equation as a compatibility condition for a linear isospectral problem, so
that the
initial value problem may be solved by the inverse scattering transform (IST)
method.

\noindent
{\bf The unidirectional model.}
Consider Euler's equations for an inviscid incompressible fluid of uniform
density
with one horizontal velocity component $u$
in the $x$ direction, and $w$ in the vertical ($z$) direction.
The fluid is acted on by the acceleration of gravity, $g$,
and is moving in a horizontally infinite domain with an upper free surface at
$z=\zeta(x,t)$ and flat bottom at $z=-h_0$.
Substituting the solution form motivated by shallow water asymptotics\citr{CH},
$u = u(x,t), \quad w = -(z+h_0) u_x $, into the conserved energy (kinetic +
potential)
for Euler's equations, and explicitly performing the $z$-integration leads to
the energy
$H_{GN} = \haf \ir dx \left[ \eta u^2 + { 1 \ov 3}\eta^3 u_x^2 + g (\eta-h_0)^2
\right]$,
where $\eta = \ze + h_0$ is the height of the water above the bottom.
Substituting the same solution form above into Euler's equations and
integrating over the vertical coordinate leads to the \GN (GN)
equations\citr{GN}.
The GN equations conserve the energy $H_{GN}$. In fact, they
are expressible in \Ham form\citr{DH} as
\beq
\left(
\begin{array}{c}
     m_t \\ \eta_t
\end{array}
\right)
= -
\left(
\begin{array}{cl}
  \partial m + m \p & \eta \p \\
 \partial \eta & 0
\end{array}
\right)
\left(
\begin{array}{c}
{\de H_{GN} / \de m} \\
{\de H_{GN} / \de \eta}
\end{array}
\right)
\labeleq{lph}
\eeq
where the momentum density $m$ is defined by
$m=\de H_{GN}/ \de u $.
The GN equations do not necessarily refer to a thin-domain expansion
in a small parameter $\ep$ that measures the ratio of depth to wavelength.
In such an expansion the kinetic energy of vertical motion
($ \sim \eta^3 u_x^2$)
in $ H_{GN}$ would be $ O(\ep^2)$.
Shallow water theory makes a further small-amplitude
assumption, in the form
$\eta=h_0+O(\al), \quad \al<< 1$,
and balances $\al=O(\ep^2)$.
In contrast, the \Ham $H_{GN}$ retains nondominant terms
(e.g., $\zeta^3$) that would be higher order in such an expansion.
Starting from the GN equations, further small-amplitude asymptotics and
restriction to unidirectional propagation in a frame moving near
the critical wave speed $c_0=\sqrt{gh_0}$, leads to the
KdV equation \citr{W},
$u_t+c_0u_x+ {3 / 2} \ uu_x+{ 1 / 6} \ c_0h_0^2   u_{xxx}=0$,
or, with the same order of accuracy in the thin-domain expansion,
the BBM equation\citr{BBM},
$u_t+c_0u_x+{3 / 2} \ uu_x- { 1 /6} \ h_0^2 u_{xxt}=0$.
Instead of making asymptotic expansions in the equations of motion,
as in the derivations of the KdV and BBM equations,
our approach in deriving \req{cluv} is to make a unidirectional approximation
by relating $m$ to $\eta$
in the GN system and preserving the momentum part of its \Ham structure
(\ref{lph}).
For this purpose, we will set
$ \eta= h_0 \sqrt{m /( h_0 c_0)}$,
and since $\eta \rightarrow h_0$ as $|x| \rightarrow \infty$ the boundary
conditions
on $m$ will be assumed to be $m \rightarrow h_0 c_0$ as $|x| \rightarrow
\infty$.
The functional $C=\ir \sqrt{m} \ dx$  is the Casimir for the Hamiltonian
operator
$(m\p +\p m)$ and so
we will refer to this invariant manifold as the Casimir manifold for \req{lph}.
Next, we scale $u \rightarrow \al u$ in the \Ham $H_{GN}$, look for
$m$ in the form $m=h_0c_0+\al m_1 + \al^2 m_2 + \al^3 m_3 + \dots$ and expand
$H_{GN}$
accordingly. With this scaling and expansion,
defining $m$ as the variational derivative of the \Ham with respect
to $u$, and balancing at order $O(\al^2)$ gives\citr{ndt}
$m_1= 2 (h_0 u - { h_0^3 u_{xx} / 3})$.
The \Ham  may then be rewritten as $H_{GN}=H_{1D}+O(\al^3)$, where
$H_{1D}= \al^2/4\ir m_1 u \ dx+\al/2  \ir m_1 c_0  \ dx$,
and the factor $1/2$ arises from restricting to a submanifold.\citr{Olv88}

The  $O(\al)$ equation of motion for $m$ on the Casimir manifold is therefore
$m_t=-(m\p +\p m){\de H_{1D} / \de m}=-{\al / 2} (m\p +\p m)u - c_0/2 \ m_x$,
or, in terms of $u$,
\beq
u_t-{ 1\ov 3}h_0^2 u_{xxt}+c_0u_x+{3 \ov 2}\al u u_x -{ 1\ov 6}h_0^2 c_0
u_{xxx}=
{ 1\ov 3} \al h_0^2 u_x u_{xx}+{ 1\ov 6}\al h_0^2  u u_{xxx}.
\labeleq{cluv0}
\eeq
Dropping the right-hand side of this equation
gives BBM or KdV, modulo replacing $u_{xxt}$ by $-c_0 u_{xxx}$.\citr{W}
Thus  \req{cluv0} can be seen as a BBM
equation extended by retaining higher order terms (selected by the \Ham
approach) in an asymptotic expansion in terms of the small-amplitude parameter
$\al$.
The restriction to the Casimir manifold  is equivalent at order $O(\al)$
to the unidirectionality assumption $\ze=\sqrt{h_0 / g} \ u+O(\al)$ in the
usual derivations of the
KdV  and BBM  models from the Boussinesq system\citr{W}\citr{Olv84}.
In fact,
$\ze= \sqrt{h_0 / g} \left[ u -{h_0^2 / 6}  u_{xx} \right]+ O(\al)$, and
in a thin-domain approximation the double derivative term in this expression
would acquire a factor $\ep^2$.

Rescaling \req{cluv0}, dropping $\al$, and going to a frame of reference moving
with speed $\ka=c_0 /2 $ reduces the equation to the standard form \req{cluv}.
Notice that \req{cluv}, like BBM, is not Galilean invariant, i.e.,
not invariant
under  $u \rightarrow u+\ka,\ t \rightarrow t,\ x \rightarrow x+\ka t$.  Thus,
equation (\ref{cluv}) is best seen as a member of a family of equations
parameterized by the speed
$\ka$ of the Galilean frame.

Using the identity
$(1-\p^2) e^{-|x|}=2 \de(x)$ and setting $\K[v]\equiv\ir dy
\exp({-|x-y|})v(y)$, expresses
equation \req{cluv} in nonlocal form as
$u_t+uu_x +\ka \K[u_y]=-\K[ u u_y + 1/2 \ u_y u_{yy}]$.
Dropping the quadratic terms on the right-hand side of
this equation gives the one studied by
Fornberg and Whitham\citr{FW}.
Fornberg and Whitham show that traveling wave solutions of
this truncated equation have a peaked limiting form. Moreover, nonsymmetric
initial data with two inflection points in their case can develop a vertical
slope
in finite time.

In a later paper we will discuss the parameterized family \req{cluv}.
The present paper focuses on the limiting case $\ka=0$,
\beq
u_t-u_{xxt}=-3uu_x+2u_xu_{xx}+uu_{xxx}\ ,
\labeleq{luv}
\eeq
where $u$ is defined on the real line with vanishing boundary conditions at
infinity and
such that the \Ham
$H_1= \haf \ir (u^2+u_x^2) \ dx $
is bounded.
As with \req{cluv0}, $H_1$ generates the flow \req{luv} through
$m=u-u_{xx}$,  $m_t=-(m\p+\p m) {\de H_1 / \de m}$.

\noindent
{\bf Steepening at inflection points.}
Consider an initial condition
that has an inflection point at $x = \bar{x}$, to the right of its maximum, and
 decays
to zero in each direction sufficiently rapidly for $H_1$ to be finite.  Define
the
the time dependent slope at the inflection point as $s(t)=u_x(\bar{x}(t),t)$.
Then
the nonlocal form of \req{luv} (with $\ka=0$) and standard Sobolev estimates
yield a differential inequality  for $s$,
${ds / dt}\leq -  s^2 / 2 +H_1$.
Hence, the slope becomes vertical in finite time, provided it is initially
sufficiently
negative. If the initial condition is antisymmetric, then
the inflection point at $u=0$ is fixed and ${d \bar{x}/ dt}=0$,
due to the symmetry $(u,x) \rightarrow (-u,-x)$ enjoyed by \req{luv}.
In this case, no matter how small $|s(0)|$, verticality
develops in finite time.
This steepening property implies that traveling wave solutions of \req{luv}
cannot have the usual bell shape since inflection points
may not be stationary in time. In fact the traveling wave solution is
given by $u(x,t)=c \exp({-|x-ct|})$.
This solution travels with speed $c$ and has a corner
(that is, a finite jump in its derivative)
at its peak of height $c$.\citr{ndrc}

\noindent
{\bf $N$-soliton solution.}
Motivated by the form of the traveling wave solution, we make the following
solution
ansatz for $N$ interacting peaked solutions,
$u(x,t)= \sum_{i=1}^N p_i(t) \  \exp({-|x-q_i(t)|}) $.
Substituting this into equation (\ref{luv}) yields
evolution equations for $q_j$ and $p_j$,
that are Hamilton's canonical equations, with \Ham $H_{A}$ given by
substituting the solution Ansatz above
into the integral of motion $H_1$, yielding
$H_{A}=1/2 \sum_{i,j=1}^N p_ip_j \exp({-|q_i-q_j|})$.
Hamiltonians of this form describe geodesic motion.
The peak position $q_i(t)$ is governed by geodesic motion
of a particle on an
$N$-dimensional surface with inverse metric tensor
$g^{ij}({\bm q})=\exp({-|q_i-q_j|})$, ${\bm q} \in \barr^N$.
The metric tensor is singular whenever $q_i=q_j$.

\noindent
{\bf Two-soliton Dynamics.}
Consider the scattering of two solitons that are initially well separated,
and have speeds $c_1$ and $c_2$, with $c_1>c_2$ and $ c_1>0$, so that they
collide.
The \Ham system governing this collision possesses two
constants of motion, $H_0 = p_1+p_2 = c_1+c_2$ and $H_A=(c_1^2+c_2^2)/2$.
Notice that if the peaks were to overlap, thereby producing $q_1-q_2=0$ during
a collision,
there would be a contradiction $2H_A=(c_1+c_2)^2 = c_1^2+c_2^2$, unless $p$
were to diverge
when the overlap occurred.

The solution of Hamilton's canonical equations for \Ham $H_A$
when $N=2$ is given by
\beqa
q_1-q_2 &=&-\log \left|{4(c_1-c_2)^2\ga e^{(c_1-c_2)t}
\ov (\ga e^{(c_1-c_2)t}+4c_1^2)(\ga e^{(c_1-c_2)t}+4c_2^2)} \right|\ ,
\nnq
\\[12pt]
p_1-p_2&=&\pm(c_1-c_2){\ga e^{-(c_1-c_2)t}-4c_1c_2 \ov\ga e^{-(c_1-c_2)t
}+4c_1c_2}\
\labeleqa{solpx}
\eeqa
and the conservation law for $p_1+p_2$.
Here $\ga$ is a constant specifying the initial separation of the peaks, and
$c_1$ and
$c_2$
are the asymptotic $t \rightarrow \pm \infty$ values of their speeds, or
amplitudes.
The divergence of $p_1$ and $p_2$ in equation (\ref{solpx}) associated with
soliton overlap can only
occur when $c_1$ and $c_2$ have opposite signs.
That is, only ``head-on"  collisions can lead to overlapping peaks (see
Fig. 1, available from authors,
for the ``soliton-antisoliton" case $c_1=-c_2=c$).

The two soliton solution
\req{solpx}  determines the
``phase shifts," i.e., the shifts in the asymptotic position
for $t \rightarrow \infty$,
that the solitons experience after interaction.
As $t \rightarrow +\infty$ the solitons re-emerge unscathed, the faster
(and larger) soliton ahead of the slower (and smaller) one.
Defining the phase shift for the faster soliton
to be $\Del q_f \equiv q_2(+\infty)-q_1(-\infty)$,
and for the slower soliton,
$\Del q_s \equiv q_1(+\infty)-q_2(-\infty)$,
leads to
$\Del q_f = \log \left( c_1^2 / (c_1-c_2)^2 \right)$, and
$\Del q_s = \log \left( (c_1-c_2)^2 / c_2^2 \right)$.
These formulae show that when $c_1 / c_2>2$ both solitons experience a forward
shift. For $1<c_1/c_2<2$ the faster soliton is shifted forward while
the slower soliton is shifted backward. When $c_1/c_2=2$ no shift occurs
for the slower soliton.

\noindent
{\bf BiHamiltonian structure.}
Equation \req{luv} follows, as well, from an action principle expressed in
terms of a
velocity potential. This action principle leads to an additional conserved
quantity,
$H_2=\haf \ir (u^3+uu_x^2) dx$, and {\it another} \Ham operator, $\p - \p^3$.
Our equation \req{luv} then can be written in \Ham form in two different ways,
$m_t=-(\p -\p^3){\de H_2 / \de m}=-(m\p+\p m){\de H_1 / \de m}$.
The two \Ham operators
$B_1 = \p-\p^3$, and $B_2=  \p m+m\p$ form a \Ham pair.
That is, their sum is still a Hamiltonian operator\citr{Olv86}.
Equation (\ref{luv}) is thus biHamiltonian and has
an infinite number of conservation laws recursively related to each other by
$B_1{\de H_n / \de m}=B_2{\de H_{n-1} / \de m} \equiv -m_t^{(n+1)}$, $n=0,\pm
1,\pm 2,\dots$.
Starting from $H_1$ or $H_2$ this relation generates an infinite sequence of
conservation laws including, e.g., $H_0=\ir m \ dx$, $H_{-1}=\ir \sqrt{m} \
dx=C$,
$H_{-2}= \haf\ir\left[ {m_x^2 / 4 m^{5 / 2}}-{2 / \sqrt{m}} \right] dx$, etc.
Correspondingly, the recursion operator $\R=B_2B_1^{-1}$ generates a hierarchy
of
commuting flows, defined by
$m_t^{(n+1)} = K_{n+1}[m] = \R K_n[m]$, $n=0,\pm 1,\pm 2,\dots$.
The first few flows in the hierarchy are
$m_t^{(0)}= -(\p - \p^3) (2\sqrt{m})^{-1}$, $ m_t^{(1)} = 0$,
$m_t^{(2)} = -m_x$, and $ m_t^{(3)} = - (m\p +\p m)u$. The last of these is our
equation \req{luv}
and the first is an extension of the integrable Dym equation\citr{AS}.  It
turns out that
all the flows in this hierarchy are isospectral and thus completely integrable.

\noindent
{\bf The isospectral problem.}
In order to find the isospectral problem for our equation, we
follow Gel'fand and Dorfmann\citr{GD} in considering the skew symmetric
spectral problem,
$(\la B_1-B_2)\phi=0$.
A class of solutions of this problem are related by $\phi=\psi^2$ to the
solutions $\psi$ of a second order symmetric spectral problem. By imposing
isospectrality,
$\la_t=0$, our equation \req{cluv} follows from the compatibility condition
$\psi_{xxt}=\psi_{txx}$ of the system for $\psi(x,t)$,
\beq
\psi_{xx}=\left[{1 \ov 4} - {m(x,t)+\ka  \ov 2 \la} \right] \psi, \quad \quad
\psi_t= -\left(\la+u\right)\psi_x+\haf u_x\psi.
\labeleq{spctr}
\eeq
This is the isospectral problem we seek.
The system \req{spctr} provides a means of solving  the initial value problem
for \req{cluv} by the purely linear IST technique\citr{AS}.
For instance, if the boundary conditions on $m$ are taken to be zero at $x=\pm
\infty$
(sufficiently fast),\citr{prdc} then the spectral problem \req{spctr} when
$\ka=0$ has a purely
discrete spectrum since $\psi(x) \rightarrow  \exp(\pm  x/2)$ as
$|x| \rightarrow \infty$, i.e., eigenfunctions always decay exponentially at
infinity.
If, e.g., the initial condition $u(x,0)$ is chosen such that
$u(x,0)=A \left({\pi / 2} \ e^x-2 \sinh x \arctan \left( e^x \right)-1
\right)$, so that
$m(x,0)=A \sech^2(x)$, for an arbitrary constant $A$,
then it is easy to show\citr{CW} that the eigenvalues $\la$ for \req{spctr} are
given by
$\la_n={2 A / [(2n+1)(2n+3)]}$, $n=0,1,2,\dots$.
This formula shows explicitly that $\la=0$ is an accumulation point for the
discrete spectrum and the eigenvalues converge to it as
$1/n^2$, $n\rightarrow \infty$, a fact that holds in general
for any initial condition decaying exponentially fast at infinity.
Equations \req{spctr} also imply that the $N$-soliton mechanical system
with \Ham $H_A$ is completely integrable\citr{nslt}.
When $\ka \neq 0$, i.e., for an equation in the family
\req{cluv}, the limiting behavior of $\psi$ becomes
$\psi(x) \rightarrow
       \exp \left({\pm {x \sqrt{{1 / 4}-{\ka / 2 \la}}}} \right)$ as
$|x| \rightarrow \infty$,
and so continuous spectrum develops out of the origin in the
interval $0\leq \la \leq 2 \ka$. Also, for $\ka\neq 0$ the soliton solution of
\req{cluv}
becomes $C^\infty$ and there is no derivative discontinuity at its peak.
The peculiar feature of the disappearance of continuous spectrum
in the limit $\ka \rightarrow 0$ can be traced to the
constant $1/4$ in the spectral problem \req{spctr}, which in turn is generated
by the
first derivative operator in $B_1$.

Numerical simulations\citr{CHH2} confirm the analysis discussed here and
demonstrate the robustness of the peaked soliton solutions.
These simulations clearly illustrate the inflection point mechanism
by which a localized (positive) initial
condition breaks up into a height-ordered train
of peaked solitons moving to the right, with the tallest ones ahead.

For their helpful and sometimes challenging remarks during the course of this
work,
we thank M. Ablowitz, I. Gabitov, I. M. Gel'fand, J. M. Hyman, B. Kupershmidt,
P. Lax,
C. D. Levermore, S. V. Manakov, L. Margolin, P. Olver, T. Ratiu, H. Segur and
T. Y. Wu.
This work is partially supported by the U.S. Department of Energy CHAMMP
program.
\vspace{1cm}
\footnoterule
\vspace{0.7cm}

\noindent
{\bf Figure caption.}
\small
\vspace{0.3cm}

\noindent
Fig. 1. The soliton-antisoliton solution $u$ reconstructed from equation
(\ref{solpx}) is
$u(x,t)=c [\exp(-|x-1/2 \ q(t)|)- \exp(-|x+1/2 \ q(t)|)]/ \tanh(ct)$.
This solution displays the steepening behavior
discussed in the text. The slope becomes vertical and the amplitude of the
solution
becomes (everywhere) zero right at the moment of overlap. At later times
the peaks redevelop and depart again according to the symmetry $(u,t)
\rightarrow (-u,-t)$.

\vspace{1cm}
\footnoterule


\begin{thebibliography} {99}
\bibitem{prdc}
One may also investigate the isospectral structure for \req{luv} using
\req{spctr}
in the periodic domain.
\bibitem{nslt}
Substituting the N-soliton solution of equation \req{luv} into the isospectral
problem
\req{spctr} implies
$\la \psi(q_i,t)= \sum_{j=1}^N p_j \exp({-\haf|q_i-q_j|}) \ \psi(q_j,t)$.
This expression constitutes a matrix eigenvalue problem for the
eigenvector defined by $\Psi_i(t) \equiv \psi(q_i(t),t)$,
i.e.,
$L{\bm\Psi}=\la{\bm \Psi}$,
with the matrix
$L_{ij}(t) \equiv p_j \exp(-|q_i-q_j| / 2)$.
The evolution equation for $\bm \Psi$ can be obtained directly from the time
part of \req{spctr},
namely
${d / dt}\left( L{\bm \Psi} \right)= A{\bm \Psi}$,
for a certain matrix $A$.
This equation and the eigenvalue problem for $\bm \Psi$
imply that $L$ evolves according to
$L_tL=AL-LA=[A,L]$,
which shows that constants of motion can be generated by taking traces of
powers
of $L$, since
${d / dt}(\hbox{Tr} \ L^n)=0$.
For instance, the first two constants of motion are
$\hbox{Tr} \ L=\sum_{j=1}^N p_j=H_0$ and $\hbox{Tr} \ L^2=H_A$.
Constancy of the determinant of the isospectral $N$-soliton matrix $L$
explicitly shows, in analogy with
the two soliton case, that soliton overlap ($q_i=q_j$ at some instant) can only
occur
if the corresponding momenta $p_i$, $p_j$ diverge to infinity. This divergence
when the
peaks overlap can be interpreted as an interchange of the corresponding
momenta. The $L$
eigenvalues are the asymptotic peakon velocities as $t\rightarrow \pm \infty$.
\end{thebibliography}
\end{document}